%% file: hopp.tex
\documentclass[12pt]{article}
\usepackage{latexsym}
\usepackage{epsf}

\thicklines                         
\input defs.tex

\def\rcite#1{ref.~\cite{#1}}
\def\hgf{\hat\g_5}
\def\gf{\g_5}
\def\kfr{\k}

\begin{document}
\hyphenation{fer-mio-nic per-tur-ba-tive pa-ra-me-tri-za-tion}

\noindent July 2000 \hfill TAUP--2632--00 \\
\null \hfill UTCCP-P-89

\begin{center}
\vspace{15mm}
{\large\bf Overlap-Dirac fermions with a small hopping parameter}
\\[15mm]
Maarten Golterman$^a$\ \ and \ \ Yigal Shamir$^b$
\\[10mm]
\end{center}
\begin{quotation} \noindent
{\small\it $^a$Center for Computational Physics, University of
Tsukuba, \\
Tsukuba, Ibaraki 305-8577, JAPAN \\
and\\
Department of Physics, Washington University,
St.~Louis, MO 63130, USA$^*$}\\
maarten@aapje.wustl.edu
\\[5mm]
{\small\it $^b$School of Physics and Astronomy,
Beverly and Raymond Sackler Faculty of Exact Sciences,
Tel-Aviv University, Ramat~Aviv,~69978~ISRAEL}\\
shamir@post.tau.ac.il
\\[10mm]
\end{quotation}
\begin{center}
{ABSTRACT}
\\[2mm]
\end{center}

\begin{quotation}
We consider overlap-Dirac fermions at non-zero bare coupling
and for a small hopping parameter, or, equivalently, large $|M|$
with $M$ the domain-wall height.
We prove the existence of a phase at large positive $M$ where
the abelian axial group $U_A(1)$ is a symmetry,
and the corresponding pseudo-scalar is an exact Goldstone boson.
We also provide a conjecture for the phase diagram
of asymptotically free gauge theories with overlap-Dirac fermions.
In particular, we suggest that, for large gauge coupling, the
massive-fermion phase at negative $M$ possibly extends to all $M<4$.
\end{quotation}

\vfill
${}^*$ Permanent address



\newpage

\noindent {\it 1. Introduction.}
In recent years it has become clear that the chiral limit can be separated
from the continuum limit in QCD-like lattice gauge theories~[1-15].
Domain-wall fermions~[1-3] provide a simple way of approaching this limit,
and have become popular in numerical simulations
(see \rcite{tb} for a review).
On the theoretical side, a key role in the formulation of the chiral limit
is played by the algebraic Ginsparg-Wilson relation~\cite{gw}
(see \rcite{fn} for a review). An extensively studied solution
of the Ginsparg-Wilson relation is the overlap-Dirac operator~\cite{ov1,ov2},
which may be regarded as a certain limiting case of
domain-wall fermions~\cite{lim}.

A common feature of all numerical simulations which attempt to approach the
chiral limit is that they show a pattern of increasing
chiral symmetry violations when the bare coupling is increased.
Keeping such symmetry violations under control in the most
economic way is important, and
analytic results valid beyond the weak-coupling limit
can provide valuable clues as to what is the best way of achieving
this goal.

In this paper we consider a
euclidean $SU(N_c)$ lattice gauge theory with $N_l$ copies of
overlap-Dirac fermions and with a finite bare coupling $g$.
For a small hopping parameter $\kfr$,
the theory may be in a massive or in a massless phase
(depending on whether $A=1$ or $A=-1$ in \seeq{X} below).
Our main result is that, in the massless phase, the theory may be reformulated
such that the abelian axial group $U_A(1)$ becomes a manifest symmetry
of both the action and the measure,
without spoiling the locality of the theory.
As a result, there are $N_l^2$ lattice pions in that massless phase.
The term {\it lattice pion} here denotes a pseudo-scalar state
which is an exact Goldstone boson for finite $g$.

To avoid additional complications which are besides our main point,
we mainly consider lattice theories where the limit $g\to 0$ defines
a confining continuum theory.
We then expect chiral symmetry to be spontaneously broken
in the continuum limit as well~\cite{bc}.
(Basically this means that the number of flavors is small enough
compared to $N_c$; see below for a comment on the case $N_c=3$.)
If the continuum theory has $N_f$ flavors,
the number of pions is $N_f^2-1$. The small-$\kfr$ phase referred to
above thus has one extra (lattice) pion compared to QCD with $N_f=N_l$.
This indicates some sort of species doubling,
and we will argue that, if the continuum limit is taken inside
that phase, the number of flavors is actually $N_f=16 N_l$,
as for naive fermions.

Our work was motivated by two recent strong-coupling 
calculations~\cite{IN,bs}.  The results of those papers will be
compared with ours in the last section.

This paper is organized as follows. In Sect.~2 we review some relevant
properties of the overlap-Dirac operator, discussing in particular
the physical significance of L\"uscher's axial transformations~\cite{ml,fn}.
In Sect.~3 we prove the main result, and  in Sect.~4 we comment on the role
of the index of the overlap-Dirac operator.
In Sect.~5 we discuss the continuum limit in some more detail.
In Sect.~6 we turn the existing results
into a conjecture on the phase diagram of overlap fermions.
Finally, in Sect.~7 we compare our results with those of refs.\ \cite{IN,bs},
and list some issues for future research.
Locality of the reformulated action is proved in the Appendix.

\newpage
\vspace{2ex}
\noindent {\it 2. The overlap-Dirac operator.}
The Ginsparg-Wilson relation is (we work
in units of the lattice spacing)~\cite{gw}
\beq
  D\,\gf + \gf\,D = D\, \gf \, D \,.
\label{GW}
\eeq
The overlap-Dirac operator~\cite{ov2} which satisfies the Ginsparg-Wilson
relation is defined by
\beq
  D = 1 + \gf H/|H| \,,
\label{D}
\eeq
\beq
  H = \gf X \,, \qquad
  X = A  + \kfr \left( \sum_\m \g_\m C_\m - B \right) \,,
\label{X}
\eeq
where $X$ is the usual Wilson-Dirac operator, and
\beq
   (C_\m)_{x,y}
   = \half ( \d_{x+\hat\m,y} U_{x,\m} - \d_{x-\hat\m,y} U^\dagger_{y,\m})
\,,
\label{C}
\eeq
\beq
  B_{x,y}  =  \half \sum_\m
  (\d_{x+\hat\m,y} U_{x,\m} + \d_{x-\hat\m,y} U^\dagger_{y,\m}) \,.
\label{B}
\eeq
Since $D$ is unchanged if $H$ is multiplied
by an arbitrary positive number, we use this freedom to set
$A=\pm 1$ in \seeq{X}.
The hopping parameter $\kfr$ is positive by convention.
(A sign flip of $\kfr$ can be undone by the transformation
$\j_x \to (-1)^{x_1+x_2+x_3+x_4}\,\j_x$ and similarly for
$\bar\j_x$.) We take the Wilson parameter to be $r=1$,
but the discussion can easily be generalized to other values.

The action $S(\j,\bar\j) = \sum \bar\j \, D \, \j$
is invariant under L\"uscher's gauge-field dependent axial
transformation~\cite{ml,fn}
\bqry
  \d\j & = & T \hgf\,\j\,, \qquad \hgf=\gf(1-D) = -H/|H| \,,
\NON
  \d\bar\j & = &  \bar\j\,T \gf \,.
\label{L}
\eqry
Here $T$ is a $U(N_l)$ generator acting on the flavor indices.
Since $(\hgf)^2=1$,  L\"uscher's axial transformations together
with the usual vector transformations generate a
$U_L(N_l)\times U_R(N_l)
=U_V(1)\times U_A(1)\times SU_L(N_l)\times SU_R(N_l)$
symmetry of the lattice action.
This statement is true for any $\kfr$.
To avoid confusion we recall that if $\kfr=O(1)$ the fermion measure
is in general not invariant under L\"uscher's $U_A(1)$,
leading to the expected axial anomaly~\cite{ml} and a massive $\h'$  particle.

The symmetry\seneq{L} does not always have
the physical significance of an axial symmetry.
Let us first recall the quark spectrum described by a single
overlap-Dirac field. As usual, the quark spectrum is determined by
the free theory. We then have for $B_0$ and $C_0$, in momentum space,
\beq
C_{0\mu}(p)=i\sin{p_\mu}\,,\ \ \ \ \ B_0(p)=\sum_\mu\cos{p_\mu}\,.
\label{BCMOM}
\eeq
It is easy to check that at the corners $p_c$ of the Brillouin zone
($p_{c\mu}=0$ or $\p$, all $\m$),
and only there, $D_0(p_c)$ is equal to either 0 or 2.
A given corner of the Brillouin zone gives rise to a massless quark field
in the continuum limit if and only if $D_0(p_c)=0$.

In the context of domain-wall or overlap-Dirac fermions,
the customary parametrization of the Wilson-Dirac operator in
\seeq{X} is
\beq
  X = 4-M  + \sum_\m \g_\m C_\m - B \,.
\label{XM}
\eeq
Comparing \seeqs{X} and\seneq{XM}, we see that
$\kappa=1/|4-M|$, and that $A=1$ ($A=-1$) corresponds to $M<4$ ($M>4$).
When $n$ components of the four-momentum are equal to $\p$
and the rest are zero, one has $B_0(p_c)=4-2n$.
Referring to the parametrization of \seeq{XM},
it follows that for $M<0$ all corners have $D_0(p_c)=2$,
while for $M>8$ all corners have $D_0(p_c)=0$.
For $M<0$ there are no massless quarks,
and when $M$ is increased above the values $0,2,4,6$ and $8$,
the numbers of massless quarks that are {\it added} to the spectrum
are $1,4,6,4$ and $1$ respectively. For $M>8$ there are 16 massless quarks.
Note that the points $M=0,2,\ldots,8,$
represent discontinuities in the spectrum.

Suppose now that $D_0(p_c)=0$ at some corner of the Brillouin
zone. Near that corner L\"uscher's transformation\seneq{L} reduces to an
ordinary axial transformation (cf.\ \seeq{dq} below) when acting on the
corresponding massless-quark state.
We recall that the physical axial charge is equal or opposite to
the lattice axial charge depending on whether
$n$ is even or odd, respectively~\cite{ks}.

The other possibility is that $D_0(p_c)=2$ at a corner.
In this case, \seeq{L} reduces to $\d\j = - T \gf\,\j$,
$\d\bar\j = \bar\j\,T \gf$. Because of the minus sign in the
$\j$ transformation rule, this is no longer an axial transformation.
There exists another version of L\"uscher's transformations where
the $\bar\j$ and $\j$ rules look more symmetric,
given by~\cite{ml}
$\d\j = T \gf(1-\half D)\j$, $\d\bar\j = \bar\j\,T (1-\half D)\gf$.
In this form, the transformation still reduces to an ordinary
axial transformation when $D_0(p_c)=0$. But for $D_0(p_c)=2$ this becomes
$\d\j =\d\bar\j =0$. In other words, for $D_0(p_c)=2$
the transformation does not act at all
on states with a momentum close to $p_c$.

One can summarize the situation
by saying that, unlike an ordinary axial symmetry,
L\"uscher's symmetry by itself does not imply the existence of
massless quarks. But if massless quarks exist, it acts on them
as an ordinary axial symmetry.
In fact, these requirements single out the transformations\seneq{L}
almost uniquely. For consider taking
$\delta\psi=T\gamma_5(1-{\cal O})\psi$
for some ${\cal O}$, with ${\cal O}=0$ for $p=0$, but possibly non-zero
for other momenta in order to accommodate the removal of doublers.  We also
want to form the same Lie algebra as in the continuum, 
so we need $(\gamma_5(1-{\cal O}))^2=1$.
From this it follows immediately that ${\cal O}$ satisfies the
Ginsparg-Wilson relation.

\vspace{2ex}
\noindent {\it 3. The massless small-$\kfr$ phase.}
The properties of $D$ for a small hopping parameter $\kfr$
rest on the fact that $X=\pm 1+O(\kfr)$ (cf.\ \seeq{X}).
It follows immediately~\cite{IN}
that $D=2-O(\kfr)$ for $A=1$ while $D=O(\kfr)$ for $A=-1$
(these statements refer to the eigenvalues of $D$;
we recall that $|D| \le 2$ always, see \rcite{ov2}).
In the case $A=1$, since $D=2-O(\kfr)$ regardless
of $g$, all correlation lengths in
the fermion sector must be finite in lattice units,
and moreover tend to zero for $\kfr\to 0$.
Hence $A=1$ corresponds to a massive-fermion phase.
(This is the usual situation in a hopping expansion.
It can be established using the methods of \rcite{vw},
or, more explicitly, by applying the techniques of the Appendix
to obtain bounds on the kernel of $D^{-1}$ for $A=1$ and small $\kfr$.)

In this section we will show that
$A=-1$ corresponds to a massless phase with $N_l^2$ lattice pions.
The first important observation~\cite{IN} is that the leading, $O(\kfr)$,
term in the expansion of $D$ is proportional to the naive-fermion operator.
Also, we have already seen that for small $\kfr$ (large $M$)
overlap-Dirac fermions undergo the same, maximal,
doubling in the continuum limit as naive fermions.
(However, overlap-Dirac
and naive fermions have a different massless spectrum
(of gauge invariant states)
at finite $\kfr$ and $g$, see below. Note that
the $\kfr \to 0$ limit is singular for normalized expectation values:
in this limit there are massless fermions, whereas setting $\kfr=0$
gives rise to no propagation at all. The hopping expansion in the
$A=-1$ phase is thus qualitatively different from that for Wilson
fermions, where, as in the $A=1$ phase, it is an expansion around
an infinite fermion-mass theory.)

In order to analyze the situation for a small but finite hopping parameter
it is convenient to introduce new variables~\cite{IN}
\beq
  q=(2-D)\,\j \,, \qquad \bar{q}=\bar\j \,.
\label{q}
\eeq
In terms of the new variables, L\"uscher's transformation\seneq{L}
reduces to the ordinary axial transformation
\beq
  \d q  = T \gf\, q\,, \qquad \d\bar{q} = \bar{q}\, T \gf \,.
\label{dq}
\eeq
The partition function is rewritten as
\bqry
  Z = {\rm det}(D) & = & \int \cd\j\cd\bar\j \, \exp(-S(\j,\bar\j))
\NON
    & = & \int \cd q \cd\bar{q} \; {\rm det}(2-D)\;
                             \exp(-S'(q,\bar{q}))\;,
\label{Z}
\eqry
where
\beq
   S'(q,\bar{q}) = \sum \bar{q}\, D(2-D)^{-1}\, q \,.
\label{Sq}
\eeq
It is easily verified that $D(2-D)^{-1}$ anticommutes with $\gf$,
as it must, in view of \seeq{dq}.
The action $S'(q,\bar{q})$ is thus invariant under ordinary
axial transformations, as well as under the usual hyper-cubic rotations.

The transition to the new variables cannot be done
for arbitrary $\kfr$ and $A$. When $\kfr$ exceeds some critical value,
$(2-D)^{-1}$ becomes singular, and in the free fermion limit
there are poles in the action  $S'(q,\bar{q})$.
Moreover, in non-trivial topological sectors $S'(q,\bar{q})$ is undefined
since there exist stable eigenmodes with eigenvalues 0 and 2.
Finally, when $A=1$, poles appear in $S'(q,\bar{q})$ already in
the hopping expansion,
and therefore the discussion below is not applicable in the massive phase.

When $A=-1$, the transition to the new variables
turns out to be a powerful tool, since in this case
\seeq{Sq} is a local action for small enough $\kfr$.
In more detail, for a range  $\kfr<\kfr_0$ of the hopping parameter
$D(2-D)^{-1}$ is bounded and has an exponentially decaying kernel
(as is the standard practice in this context, the last statement
defines the notion of locality used in this paper).
Boundedness is obvious since $D=O(\kfr)$,
while locality of $D(2-D)^{-1}$ is established in the Appendix
for $\kfr < \k_0$
(we note that the ``admissibility" constraint on the gauge field~\cite{hjl}
is not necessary for small $\kfr$).
As for the factor ${\rm det}(2-D)$ in \seeq{Z}, 
it is not expected to change the universality class
(an argument similar to that in the appendix shows that
${\rm tr}\log(2-D)$ is local).

Thus, we find that for $\kfr < \k_0$ the theory can be
consistently formulated in terms of a local action $S'(q,\bar{q})$,
where the action and, obviously, the measure are invariant under
ordinary vector and axial transformations. The fact that we are dealing
here with the simple axial transformation of \seeq{dq} is important.
As mentioned in the introduction, we restrict the discussion to
those cases where the limit $g\to 0$ defines a confining continuum theory.
We then have confinement for any $g$, and the
standard arguments that confinement implies chiral symmetry
breaking apply~\cite{bc}. In the present context, the formation
of a $\svev{\bar{q}q}$  condensate was confirmed in the strong coupling
limit in a $1/N_c$ expansion~\cite{IN} (see also \rcite{colwit}).

Therefore, the lattice theory has $N_l^2$ axial generators,
and the pseudo-scalar Goldstone bosons must be in one-to-one
correspondence with those.
We thus conclude that there are  $N_l^2$ lattice pions for $\kfr < \k_0$
and any (finite) $g$.
The (confining) small-$\kfr$ massless phase may actually
be {\it defined} as the
part of the phase diagram with $N_l^2$ lattice pions.
We expect this phase to extend beyond the region $\kfr < \k_0$
and, in fact, beyond the region where $(2-D)^{-1}$ is bounded, see below.
We comment in passing that in the case
of naive fermions the lattice symmetry is bigger~\cite{kws},
and hence the number of lattice pions is larger as well.
This naive-fermion symmetry requires that the action
has only odd-neighbor couplings.

In summary, we have established the existence of a class of
overlap-Dirac theories where for $A=-1$ and small $\kfr$ there
is only one phase for all values of $g$. This phase is characterized by
confinement and chiral symmetry breaking, and has $N_l^2$ lattice pions.

\vspace{2ex}
\noindent {\it 4. The index and $U_A(1)$.}
The question of whether $U_A(1)$ is a symmetry or not may be approached
from another direction. The integrated Ward identity
associated with a $U_A(1)$ transformation reads~\cite{ml}
\beq
  \svev{\d \co} = \svev{ {\rm Tr}(\hgf)\,\co} \,,
\label{ind}
\eeq
where $\co=\co(\j,\bar\j,U_{\m})$ and the axial variation denoted by $\d$
is defined in \seeq{L}.
(To make the above statement well defined
we may assume that we work in a finite volume.)
It is easy to see that ${\rm Tr}(\hgf)=0$ for small
$\kfr$ for all gauge field configurations, regardless of
the sign of $A$ in \seeq{X}.
First, ${\rm Tr}(\hgf)$ is (proportional to) the index of $D$ \cite{ml}.
Then, a non-zero index requires the simultaneous existence of eigenvectors
with eigenvalues zero and two~\cite{fn,hf,chiu}.
But, for $A=-1$, $D=O(\kfr)$ and there can be no eigenvalue equal to two.
For $A=1$, $D=2-O(\kfr)$ and there can be no zero eigenvalue.

What we learn is that the small-$\kfr$ global symmetry of both
the $A=1$ and the $A=-1$ phases is $U_L(N_l)\times U_R(N_l)$.
However, as already explained above,
there is an important difference between the two phases.
In the $A=-1$ phase the conserved axial generators may be taken
to be the ordinary axial generators associated with \seeq{dq}.
We have seen that this implies the existence of massless quarks
and (if there is confinement) spontaneous chiral symmetry breaking.
In contrast, in the $A=1$ phase we have only  L\"uscher's symmetry
at our disposal, and as explained in Sect.~2, this is {\it not} 
incompatible with
the absence of massless-fermion states.

\vspace{2ex}
\noindent {\it 5. Relation to the continuum limit.}
The previous results provide valuable information
for the task of mapping out the $(M,g)$ phase diagram of overlap-Dirac
fermions. As already mentioned, the hopping expansion
is an expansion in $1/|4-M|$. The  massless ($A=-1$) phase
at small $\kfr$ corresponds to $M>M_{c+}(g)$,
while the massive ($A=1$) phase corresponds to $M<M_{c-}(g)$.

We now wish to determine the end points of these two critical lines
in the continuum limit.
We will argue that the massless phase at large positive $M$
extends  down to $M=8$ for $g\to 0$, whereas
the (massive-fermion) phase at large negative $M$
extends up to $M=0$ provided $N_l \ge 2$.
The region $O(g)<M<8-O(g)$ at small $g$
is filled with a phase supporting $N_l^2-1$ lattice pions.

Let us begin with the last statement. For $0<M<8$ there are massless quarks,
and since L\"uscher's symmetry\seneq{L} acts on those as an axial symmetry,
spontaneous symmetry breaking should occur, and the corresponding
Goldstone bosons should exist. In this range no massless state
in the $U_A(1)$ channel is expected~\cite{wv,chan} 
because the index of $D$ can be non-zero. 
Since there are $N_l^2-1$ conserved
axial generators for non-zero $g$,
this must also be the number of lattice pions.
A corollary is that the massless large-$M$ phase with  $N_l^2$ lattice pions
cannot end at any $M<8$ for $g\to 0$.
Recall that, according to the free-field (or weak-coupling) analysis,
the largest value where a phase transition takes place in the $g\to 0$ limit
is $M=8$. Assuming that that analysis exhausts all possible phase transition
points for  $g\to 0$, it follows that the large-$M$ phase
must end at $M=8$.
Similarly, the massive-fermion phase with no lattice pions
at negative $M$ (which is actually a pure glue phase)
should end at $M=0$ for $g\to 0$.
The last statement is true except for
$N_l=1$, where both the negative-$M$ and intermediate-$M$ regions
do not support lattice pions, and therefore they may be analytically
connected at non-zero $g$.

For small $g$ and $M>8$ there are massless states at all the corners of the
Brillouin zone, namely 16 quark fields per each overlap-Dirac fermion.
The lattice $U_A(1)$ symmetry corresponds to a flavor non-diagonal
axial symmetry in the continuum limit~\cite{ks} (see 
Sect.~2). The total number of pions in the continuum limit
is $(16 N_l)^2 -1$, whereas for finite $g$
there are only $N_l^2$ pions.
The difference is explained by states which are approximate Goldstone
bosons for small $g$, and which become massless only
in the continuum limit. This bears some resemblance to staggered
fermions, where only one pion is exactly massless at finite lattice spacing,
while the rest become massless only in the continuum limit.

If one relaxes the assumption we have made in the introduction
about the fermion content,
one can find cases where asymptotic freedom is lost and/or
there is no chiral symmetry breaking for small $g$.
In such cases there has to be (at least) one phase transition
as a function of $g$ even at large (positive) $M$.
For the physical case $N_c=3$,
asymptotic freedom at large $M$ is lost for $N_l=2$ ($N_f=32$),
and it is not clear if there is (confinement and) chiral symmetry breaking
even for $N_l=1$ ($N_f=16$) in the continuum limit.
At strong coupling one always has confinement, and we expect our conclusion
to hold on the large-$g$ side of such a transition.
Also, whatever replaces confinement and
chiral symmetry breaking at small $g$
should still be consistent with having $16N_l^2$ massless fermions for $g=0$.
Note that this issue does not affect the phase diagram in the region
of interest for Lattice QCD, with $0<M<2$.

\vspace{2ex}
\noindent {\it 6. Conjectured phase diagram of overlap-Dirac fermions.}
In this section we put forward a conjecture for the
$(M,g)$ phase diagram of overlap-Dirac fermions.
We have already discussed three out of four boundaries of the phase diagram.
We invoke a mean-field argument to cover
the last, $g=\infty$, boundary, and then consider the
simplest finite-$(M,g)$ interpolation.

\begin{figure}[hbt]
\vspace*{0.4cm}
\centerline{
\epsfxsize=8.0cm
\put(-100,0){\epsfbox{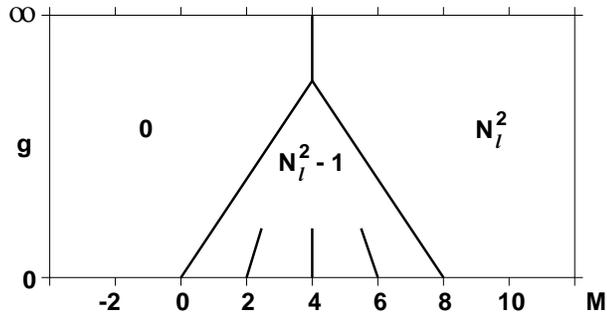}}
}
\vspace*{0.4cm}
\caption{ \noindent {\it Conjectured phase diagram of overlap-Dirac fermions
for $N_l \ge 2$.
The number of lattice pions in each phase is indicated.
}}
\label{GWfig}
\vspace*{0.5cm}
\end{figure}

The conjectured phase diagram is depicted in Fig.~1 for $N_l \ge 2$.
It contains the two phases with zero and $N_l^2$ lattice pions
whose existence was established
for large-negative and large-positive $M$ respectively.
There is also a single phase at intermediate
values of $M$, characterized by the existence of $N_l^2-1$ lattice pions.
The phase transition lines emanating from the points $M=2,4,6$ must
be there because the quark spectrum in the continuum limit
changes at those points. As explained earlier, the regions
surrounding these points at small $g>0$ all support $N_l^2-1$ lattice pions.
In our view the most plausible scenario is that
these regions are analytically connected at non-zero $g$, but this does
not have to be the case.  The phase transitions are discontinuous
at $g=0$, and therefore they are discontinuous for small non-zero $g$, by
continuity.  We expect them to be discontinuous for all $g$.
We remind the reader that this phase diagram is applicable provided
there is a confining continuum theory in the limit $g\to 0$
for all $M$.

If $D$ satisfies the Ginsparg-Wilson relation,
so does $2-D$. The passage from $D$ to $2-D$ is effected by
$X\to -X$, and the phase diagram for
$2-D$ is obtained by the replacement $M \to 8-M$.
(In the $D$-phase diagram, however, the critical lines may
not transform into each other under $M \to 8-M$,
and in that sense Fig.~1 may be misleading.)

The symmetries of the three phases are the following.
(Below $G\to H$ denotes that the global symmetry
is $G$ and that $H$ is the symmetry of the vacuum.)
The phase with $N_l^2$ lattice pions is characterized by
$U_L(N_l)\times U_R(N_l) \to U_V(N_l)$.
The phase with $N_l^2-1$ lattice pion is characterized by
$U_V(1)\times SU_L(N_l)\times SU_R(N_l) \to U_V(N_l)$.
Finally, the phase with no lattice pions has the global symmetry
$U_L(N_l)\times U_R(N_l)$ but no massless fermions, and no
spontaneous symmetry breaking.

As can be seen from Fig.~1, we believe that the two small-$\kfr$ phases
meet at $M=4$ for $g=\infty$. If this is true,
the phase with $N_l^2-1$ lattice pion
does not extend to $g=\infty$.
(It could be that the triple point is at $g=\infty$.
Note that one is usually interested
in taking the continuum limit inside the middle phase
because only in the interval $0<M<2$ there is one quark per one
overlap fermion in the continuum limit.)
Let us define $\svev{U_{x\m}}$ in a gauge-invariant way, for instance
as the fourth root of the average plaquette. Then we have that
$\svev{U_{x\m}}\to 0$ for $g\to\infty$ and,
if a mean-field analysis  is reliable,
the $g\to\infty$ limit should behave like $\kfr \to 0$.
For $M>4$ this means that we are in the massless small-$\kfr$ phase with
$A=-1$, which has $N_l^2$ lattice pions,
and for $M<4$ we are in the $A=1$ massive
phase, with no lattice pions.

If correct, the phase diagram of Fig.~1 leads to an important observation.
For $0<M<4$, as we increase $g$
at fixed $M$, we eventually move into the massive phase, thus loosing
all the massless quarks (even though the action is still invariant
under L\"uscher's symmetry)! Remembering the connection with
domain-wall fermions \cite{lim}, this means that under the same conditions
domain-wall fermions will support (no massless quarks and)
no massless pions even if the extent of the fifth dimension
is taken to be arbitrarily large.

The phase diagram of Wilson fermions is known to
contain the Aoki phase~\cite{aoki}, where parity and vector-like
symmetries are broken by a flavor non-singlet pseudo-scalar condensate.
The question may arise whether a similar phenomenon can take place
in the present context as well. The answer is no.
As explained in \rcite{ssrs}, the Aoki phase is related
to those terms in the effective chiral lagrangian
(for finite lattice spacing) that break axial symmetries explicitly.
There are no such terms in the present
situation since one has a chiral symmetry at finite lattice spacing as well.
Using the lattice chiral symmetries one can rotate
any non-singlet pseudo-scalar condensate into the usual singlet
scalar one $\svev{\bar{q}q}$.
An Aoki phase may, however, arise if an explicit quark mass term
is added to the action. This was investigated numerically in
the context of domain-wall fermions in \rcite{aokietal}.

\vspace{2ex}
\noindent {\it 7. Discussion.}
Finally, we compare our results with previous work.
Our results are in agreement with those of Ichinose and Nagao
who studied the massless phase
to second order in the hopping-parameter expansion~\cite{IN}.
The fate of the $U_A(1)$ pseudo-scalar state was left open
in \rcite{IN}, and it was conjectured that it may eventually
pick up a non-zero mass (in a higher order in $1/N_c$).
As we showed rigorously in this paper,
in fact $U_A(1)$ is an exact symmetry for a small hopping parameter,
and the corresponding pseudo-scalar is an exact
Goldstone boson for any value of $g$.
This result is consistent with the observation that,
for small $\kfr$, the lattice $U_A(1)$
symmetry becomes a flavor non-diagonal axial symmetry in the
continuum limit.

The hamiltonian strong-coupling analysis of Brower and Svetitsky~\cite{bs}
was done for domain-wall fermions with a continuous fifth coordinate
($a_5=0$) in the limit $L_5\to\infty$, where $L_5$ is the size of
the fifth dimension. We recall that in this double limit
domain-wall fermions are expected
to reduce to overlap-Dirac fermions~\cite{lim}.
This work is not limited to a small hopping parameter.
At zeroth order in the expansion in $1/g^2$
they find a massive phase for $M<3$ and a gap-less phase for $M>3$.
This leading-order result supports the mean-field argument of Sect.~6
that there is only one critical point at $g=\infty$.
(At that order there are no spatial couplings,
and for each site the problem reduces to a free one-dimensional
hamiltonian acting on the fifth coordinate.
While the critical value in \rcite{bs} is $M=3$,
rather than $M=4$, this is merely a technical difference stemming from
the fact that time is taken to be continuous in \rcite{bs}.)

The order-$1/g^2$ result of \rcite{bs}
is summarized, for $M>3$, by an effective low-energy hamiltonian
in $3+1$ dimensions which contains two terms, and (in their notation) reads
$H^{\rm eff}_{\rm s-s}+H^{\rm eff}_{\rm site}$.
The symmetry of $H^{\rm eff}_{\rm s-s}$ is that of naive fermions.
Since, at that order, $H^{\rm eff}_{\rm s-s}$ 
has only nearest-neighbor couplings,
this can be explained by the fact that
the $O(\kfr)$ term in the overlap-Dirac operator
is proportional to the naive fermion one.
At order $1/g^4$, $H^{\rm eff}_{\rm s-s}$ will contain also
next-to-nearest neighbor couplings, and
this may reduce the global symmetry to $U_L(N_l)\times U_R(N_l)$.
If we would ignore the second term, $H^{\rm eff}_{\rm site}$,
this would be in agreement with our results.

The second term in the effective hamiltonian, $H^{\rm eff}_{\rm site}$,
explicitly breaks all axial symmetries, having
the same structure as an explicit mass term for the quarks.
Although mathematically there is no direct conflict between
this (hamiltonian) result and our (euclidean) result,
physically the two results seem to be in conflict
if both $g$ and $M$ are large (and therefore both results should be valid).
We hope to resolve this issue in the future.
To this end, it may be useful to carry out a euclidean strong-coupling
analysis, for example using the method of \rcite{kws}.

In conclusion, in this paper we have analyzed overlap-Dirac fermions
with a small hopping parameter. While there are issues that require
further work, a concrete picture of the
phase diagram is beginning to emerge. In the future we hope
to generalize the discussion to domain-wall fermions
(with a non-zero $a_5$ and a finite $L_5$),
thus making closer contact with present-day
numerical simulations.  However, as we argued here, already in the
limit where L\"uscher's chiral symmetry is exact,
it is possible that a phase without
massless quarks exist at fixed $0<M<4$ and large gauge coupling.

\vspace{2ex}
\noindent {\it Acknowledgements.}
We wish to thank Mike Creutz, Ikuo Ichinose, Ben Svetitsky
and Marvin Weinstein for helpful discussions.  MG would like
to thank the Institute for Nuclear Theory at the University of
Washington for its hospitality.
This research is supported by
the United-States -- Israel Binational Science Foundation,
and Y.S. is supported in part by the Israel Science Foundation,
while MG is supported in part by the US Department of Energy.

\vspace{2ex}
\noindent {\it Appendix. Exponential localization. }
The Legendre expansion of $(H^2)^{-1/2}$ (cf.\ \seeq{X})
was used in \rcite{hjl} to prove the exponential localization of $D$.
This expansion is very informative,
and may be used to obtain exponential-localization bounds on functions of $D$
as well, in particular on $D(2-D)^{-1}$.
The Legendre expansion is given by~\cite{hjl}
\beq
  (H^2)^{-1/2} = c \sum_{k\ge 0} t^k \f_k(Z) \,.
\label{H2}
\eeq
Here $\f_k$ are Legendre polynomials, normalized such that
${\rm max}\,\f_k(z)=1$ for $-1\le z\le 1$.
With slight adaptation the other ingredients are
defined as follows. Let $v_0={\rm max}(H^2)$, $u_0={\rm min}(H^2)$,
where extremization is done
over the entire gauge-field configuration space.
We take a (positive and) small enough hopping parameter $\kfr$ so that $u_0>0$.
Then
\beq
  Z = {u_0+v_0-2H^2\over v_0-u_0} \,,
\eeq
\beq
  \half(t+t^{-1}) = {v_0+u_0\over v_0-u_0} \,,
\label{t}
\eeq
with $0<t<1$ and
\beq
  c = \left({4t\over v_0-u_0}\right)^{1/2} \,.
\eeq
Since $|B| \le 4$ and (for $A=-1$)
\beq
  H^2 = 1 + 2\kfr B + O(\kfr^2) \,,
\eeq
we have $v_0=1 + 8\kfr  + O(\kfr^2)$,
$u_0=1 - 8\kfr  + O(\kfr^2)$, $t=4\kfr + O(\kfr^2)$
and $c=1 + O(\kfr)$. We may then write (for $A=-1$)
\beq
  -D = (c-1) + c \sum_{k\ge 1} t^k
     \Big( \f_k(Z) - b\, Y \, \f_{k-1}(Z) \Big) ,
\label{Dexp}
\eeq
where $b=\kfr/t$ and $Y=\sum_\m \g_\m C_\m - B$, cf.\ \seeq{X}.
We comment that the Legendre expansion may be set up using any
$0<u\le u_0$ and $v\ge v_0$.
The ``best'' values defined above, $u_0$ and $v_0$,
guarantee that $t/\kfr=O(1)$. This is a natural relation because,
as described below, $t$ effectively plays the role of a hopping parameter.

We now turn to the operator
\beq
  D(2-D)^{-1} = \sum_{n\ge 1} (D/2)^n \,.
\label{geom}
\eeq
In order to obtain a bound
on the kernel corresponding to \seeq{geom} it is useful to
regard $t$ as an independent parameter. Doing so, we obtain the expansion
\beq
  D(2-D)^{-1} = \sum_{k\ge 0} t^k \cd_k \,,
\label{DDexp}
\eeq
where $\cd_k$ is defined by substituting \seeq{Dexp} into \seeq{geom}
and collecting all terms involving an explicit factor of $t^k$.
At order $t^k$, the operator encountered  in the expansion of $D$
(\seeq{Dexp}) allows for at most $2k$ hoppings~\cite{hjl}.
The same statement applies to the new kernels:
$\cd_k(x,y)=0$ for $|x-y|>2k$ where
$|x-y|=\sum_\m |x_\m-y_\m|$ (the ``taxi-driver distance'').
Therefore
\beq
  D(2-D)^{-1}(x,y) = \sum_{2k\ge |x-y|} t^k \cd_k(x,y) \,.
\label{xy}
\eeq
What is still needed is a bound on $\cd_k$.
We first observe that
\beq
  |\f_k(Z) - b\, Y \, \f_{k-1}(Z)| \le 1 + 8b \,.
\label{basic}
\eeq
In view of this bound we are lead to consider
\beq
  E = d + c (1+8b) \sum_{k\ge 1} t^k \,,
\label{Eexp}
\eeq
where $d=|c-1|$, as well as
\beq
  E(2-E)^{-1} = \sum_{n\ge 1} (E/2)^n =
  \sum_{k\ge 0} t^k \ce_k \,.
\label{Egeom}
\eeq
Again $\ce_k$ is defined
by collecting all terms involving $t^k$ using \seeq{Eexp}.
The double geometric series is easily summed, giving
\beq
  E(2-E)^{-1} = { d + (c+8bc-d)t \over 2-d -(2+c+8bc-d)t} \,,
\label{En}
\eeq
and an explicit expression for $\ce_k$ follows by re-expanding
\seeq{En} in powers of $t$.

Following the construction we see that, thanks to inequality\seneq{basic},
each term in $\cd_k$ is bounded by a corresponding
term in $\ce_k$. The latter is obtained if every factor of
$\f_k(Z) - b\, Y \, \f_{k-1}(Z)$ is replaced by $1 + 8b$ and
every factor of $c-1$ by $d$.
Using \seeq{xy} we thus obtain a bound (for $|x-y|\ge 2$)
\bqry
  |D(2-D)^{-1}(x,y)| & \le & \sum_{2k\ge |x-y|} t^k \ce_k \; =
\NON
& = & {d\,s^{ [\,|x-y|/2] }
      + t(c+8bc-d)s^{ [\,|x-y|/2-1] } \over  2-d -(2+c+8bc-d)t } \,,
\label{bound}
\eqry
where $s=(2+c+8bc-d)t/(2-d)$
and $[l]$ is the smallest integer $n$ such that $n \ge l$.
Recalling the value of $t=4\kfr+O(\kfr^2)$
(cf.\ \seeq{t}), the desired exponential bound
on the kernel of $D(2-D)^{-1}$ is thus given by
\seeq{bound} provided we choose $\kfr<\k_0$, where $\k_0$ is
the smallest value where the denominator in \seeq{bound} is zero.
The denominator is strictly positive for $0<\kfr<\k_0$
since both $t$ and $d$ are $O(\kfr)$.


\vspace{5ex}
\centerline{\rule{5cm}{.3mm}}

\end{document}

%% file: defs.tex
\def\d{\delta}
\def\f{\phi}                    
\def\g{\gamma}
\def\h{\eta}

\def\j{\psi}
\def\k{\kappa}

\def\m{\mu}

\def\p{\pi}                     

\def\cd{{\cal D}}
\def\ce{{\cal E}}

\def\co{{\cal O}}

\def\cbo{{\,\raise-.15ex\Sc [\,}}                       

\def\svev#1{\left\langle #1\right\rangle}       

\def\ddt#1{{\buildrel {\hbox{\LARGE .\kern-2pt.}} \over {#1}}}

\def\beq{\begin{equation}}
\def\eeq{\end{equation}}
\def\bqry{\begin{eqnarray}}
\def\eqry{\end{eqnarray}}

\def\beqn#1{ \renewcommand{\theequation}{#1}
             \begin{eqnarray} }
\def\eeqn{ \renewcommand{\theequation}{\arabic{equation}}
           \end{eqnarray} }

\def\beqr#1{ \setcounter{equation}{#1}
             \begin{eqnarray} }

\def\eeqr{\end{eqnarray}}
\def\NON{\nonumber\\}
\def\beqrabc#1{ \setcounter{equation}{0}
                \renewcommand{\theequation}{#1\alph{equation}}
                \begin{eqnarray} }
\def\beqrn#1#2{ \setcounter{equation}{#2}
                \renewcommand{\theequation}{#1.\arabic{equation}}
                \begin{eqnarray} }
\def\seeq#1{eq.~(\ref{#1})}

\def\seeqs#1{eqs.~(\ref{#1})}

\def\seneq#1{~(\ref{#1})}

\def\NPB#1{Nucl. Phys. {\bf B#1}}
\def\NPBP#1{Nucl. Phys. (Proc. Suppl.) {\bf B#1}}
\def\PLB#1{Phys. Lett. {\bf B#1}}
\def\PRD#1{Phys. Rev. {\bf D#1}}

\def\PRL#1{Phys. Rev. Lett. {\bf #1}}

\def\sstyle{\scriptstyle}

\def\frac#1#2{ {\sstyle {#1\over #2} } }

\def\half{{1\over 2}}
